# Real Time Communication Capacity for Data Delivery in Wireless Sensor Networks

Deepali Virmani[1] and Satbir Jain[2]

[1] Department of CSE, BPIT, GGSIPU, Delhi, India, 110085.
*deepalivirmani@gmail.com*

[2] Department of CSE, NSIT, DU, Delhi, India, 110003.
*Jain_satbir@yahoo.com*

**Abstract**

Real-time applications are performance critical applications that require bounded service latency. In multi-hop wireless ad-hoc and sensor networks, communication delays are dominant over processing delays. Therefore, to enable real-time applications in such networks, the communication latency must be bounded. In this paper, we derive expressions of real-time capacity that characterize the ability of a network to deliver data on time as well as develop network protocols that achieve this capacity. Real-time capacity expressions are obtained and analyzed for the earliest deadline first, deadline monotonic. This paper presents a treatment of the real-time capacity limits. The limits are derived for two extreme traffic topologies namely, the load balanced topology and the convergecast (i.e., many-to-one) topology. It considers DM and EDF scheduling algorithms, and discusses the implications of the capacity limit expressions.

## 1) Introduction

The performance-critical applications that require bounded delay latency are referred to as real-time applications. Those wireless sensor networks that are capable of providing bounded delay guarantees on packet delivery are referred to as real time wireless sensor networks. A multitude of problems needs to be solved to achieve the goal of supporting real-time applications in WSN. As mentioned earlier, the network uses shared (wireless) medium for communication. Therefore, one needs a distributed medium access control (MAC) protocol that is capable of providing guaranteed bandwidth over multiple hops. Bounded latency guarantee of end-to-end packet delivery is a necessary first step. The real-time support must be achieved at low power and low message overhead to limit interference and conserve energy. We define real-time capacity of a network to be its information carrying ability for given deadlines, where only those information bits that arrive at their destination within the specified deadline count. For WSN the unit of data is a packet rather than bits. If a packet does not reach its destination by the given deadline, then its contribution to the real-time capacity is 0. This is in contrast to the capacity defined in [1] and other research cited above where every transfer of bit counts. Analytic expressions for real-time capacity facilitate the process of designing a network that is guaranteed to meet specified throughput and delay requirements. The expressions describe values of a set of variables that will enable the network to meet anticipated real-time requirements. In other words, they define the feasibility region in the space of such variables. In the event of dynamically changing network, which is expected of WSN, besides planning and designing, the feasibility region allows optimization of the operation of the network. Abdelzaher et al. presented the first results on the real-time capacity limits of wireless sensor networks [2]. Expressions for the real-time capacity limits are obtained using sufficient conditions on schedulability. A class of time independent fixed priority scheduling algorithms is considered. Load-balanced and converge cast the two extreme traffic topologies are considered. In this paper, we present the results and extend them for the EDF scheduling algorithm. Yan et al. propose an energy-efficient service for providing differentiated surveillance by varying the degree of sensing coverage [3].

Cao et al. present an stochastic performance analysis of WSN for surveillance applications based on network parameters [4]. Ting Yan presented schedulability analysis of periodic flows in WSN [5]. Schmitt and Roedig use and extend the Network Calculus [6] to formulate the trade off between the buffer size requirements and power consumption with the worst case delay [7]. A suitable arrival curve for WSN is presented. Koubaa et al. extend the Sensor Network Calculus formulation of Schmitt and Roedig for a cluster based tree topology and IEEE 802.15.4/ZigBee MAC protocol [8]. The authors conclude that the bandwidth requirement grows exponentially with the depth of the tree. Their conclusion is based on the incorrect assumption that number of clusters grow exponentially with tree depth. The growth is in-fact quadratic in tree depth. To see this, the number of clusters at depth n is approximated by the ratio of the area contained within the ring of radii $nR$ and $(n-1)R$ and the area contained within one cluster $\Pi R^2$. Thus, the number of clusters at depth $n$ is approximately $2n$. Therefore, the total number of clusters in the tree up-to depth $n$ is $\sum_{i=n}^{n} 2i = n^2 + n$. Furthermore, the delay bound obtained is based on the Concatenation Theorem of Network Calculus. While the theorem has been proved for wired networks, it does not hold in its current form for wireless networks. This paper presents a treatment of the real-time capacity limits. The limits are derived for two extreme traffic topologies namely, the load balanced topology and the convergecast (i.e., many-to-one) topology. It considers DM and EDF scheduling algorithms, and discusses the implications of the capacity limit expressions

## 2) Problem Formulation

We consider a multi-hop wireless sensor network where the number of nodes is denoted by $n$. The set of nodes that can receive packets from node $x$ is denoted by $neighbourhood(x)$. The packets are denoted by $P_i$. The arrival time of a packet, $At_i$ is defined as the time when the packet arrives at the transmit queue of the originating node. Each packet has a relative deadline $D_i$ associated with it. The packet must be delivered to its destination by the time $At_i + D_i$. We associate a per-hop transmission time, $T_i$ with each packet. This transmission time is a function of raw bandwidth and characteristics of the physical medium. In terms of the effective bandwidth, $B$, $T_i$ is the ratio of the packet size and $B$. The per-hop utilization is defined as $T_i/D_i$. A packet contributes this much to the overall utilization of every node in its path from time $At_i$ to $At_i + D_i$. Therefore, if $K_x(t)$ is the set of packets that are present in the network at some given time $t$, whose deadline has not expired yet and whose path includes node , then the utilization of node $x$, denoted by $Ut_x$ is defined as

$$Ut_x(t) = \sum_{P_i \in K_x(t)} T_i/D_i \qquad (1)$$

Since $BT_i$ is the packet size, $BUt_x(t)$ gives the real-time capacity demand at node $x$. Therefore, the real-time capacity demand for the entire network is given by $B\sum_x Ut_x(t)$. If we obtain an upper bound on $Ut_x$ such that if the node's utilization does not exceed $Ut_x$ then all packets are guaranteed to meet their deadline, then no deadline miss would occur if the real-time capacity demand does not exceed $B\sum_x Ut_x(t)$. In other words,

$$RTCC = B\sum_x Ut_x(t) \qquad (2)$$

Where RTCC denotes real time communication capacity.

## 3) Assumptions

For channel access, we assume an ideal MAC protocol, as the MAC protocol transmits packets in priority order. The propagation speed in wireless medium is speed of light, and the distance scale is of the order of a few kilometers. Therefore, the propagation delay of the packets is negligible as compared to queuing and transmission delays. Accordingly, we ignore the propagation delay of packets in the following analysis. We do not assume circular radio range in the theoretical derivations. We, however, assume that the number of nodes that may be blocked due to a transmission is upper bounded, and that the bound is known. We assume that the per-hop transmission time of a packet remains constant throughout the route of the packet. We use the assumption of large number of nodes to perform approximations of fixed priority real-time capacity limit expressions.

## 4) Feasibility Conditions

We consider transmission of a packet through a sequence of nodes, which we call the packet's path. We now derive a path-specific condition on meeting end-to end deadlines as a bound on a function of utilization values. If the value of this function does not exceed the bound, then the packets transmitted along this path are guaranteed to meet their deadline. In the following analysis, we use theorems for schedulability of tasks contending for a given set of resource(s). In the context of WSN, the resource is the channel at the receiver node $x+1$. Let $neighbourhood(x)$ denote the set of nodes who contend for the transmission of node $x$. The broadcast nature of wireless medium allows only one node of this to transmit at a time. Therefore, all packets scheduled for transmission in $neighbourhood(x)$ form one single virtual queue whose total utilization is equal to

$$VQ_x = \sum_{i \in neighbourhood(x)} Ut_x \qquad (3)$$

Where $VQ_x$ is the neighborhood utilization. Let $u$ be upper bound on the size of $neighbourhood(x)$ for all $x$. Then $\sum_x VQ_x \leq u \sum_x ut_x$. Therefore, (2) can be written as

$$RTTC \geq B \Big/ u \sum_x VQ_x \qquad (4)$$

### 4.1) Corrections for priority inversion

In wireless networks, since no concurrent transmission can occur in the neighborhood of a receiver, the transmission of a packet can block another transmission up to two hops away. From the point of view of prioritized transmissions, this can lead to priority inversions of a unique kind [2]. For example, consider the case of a receiver receiving a high priority packet. No node in the neighborhood in the receiver can transmit simultaneously. If there is some lower priority packet sender at the periphery of the neighborhood that has packets for a receiver in the nearby neighborhood, then it gets blocked. If this latter transmission were the highest priority for the latter receiver, from the perspective of the receiver, this is a priority inversion. Since this phenomenon arises not due to waiting for a lower priority transmission, it is referred to as pseudo priority inversion. The effect of delays due to pseudo priority inversion decreases the real-time capacity. We quantify the effect of pseudo priority inversion as an increase in neighborhood utilization by a factor $1 \leq \alpha \, 2$. Applying this correction to the expression for the real-time capacity (5) gives,

$$RTCC \geq B \Big/ u\alpha \sum_x VQ_x \qquad (5)$$

### 4.2) Fixed Priority Scheduling

Let us consider some packet $P_a$. Let $TA_x$ denote the time between the arrival of the last bit of $P_a$ at node $x$ and departure of its last bit from the same node. In other words, $TA_x$ is the packet's delay at node $x$. The following theorem proved by Abdelzaher [10] gives the bound on the delay of a task as a function of utilization for time-independent fixed priority scheduling algorithms:

**Theorem 4.1** (The Stage Delay Theorem [2]) If task T spends time $TA_x$ at resource $x$, and $Ut_x$ is a lower bound on the maximum utilization at that hop, then:

$$TA_x = Ut_x \frac{(1 - ut_x/2)}{1 - Ut_x} D_{max} \quad (6)$$

Where $D_{max}$ is the maximum end-to-end deadline of all tasks of higher priority than T. This theorem applies to a very general class of resources, where the access to the resource is granted to only one task. Correspondingly for wireless sensor networks, from the stage delay theorem,

$$TA_x = \frac{VQ_x(1 - VQ_x/2)}{1 - VQ_x} D_{max} \quad (7)$$

In other words, if the neighborhood utilization of node $x$ does not exceed $VQ_x$, then the packet delay at that node does not exceed $TA_x$. The end-to-end bound on the delay is obtained by simply summing over all nodes:

$$\sum_{x=1}^{N} TA_x = \sum_{x=1}^{N} \frac{VQ_x(1 - VQ_x/2)}{1 - VQ_x} D_{max} \quad (8)$$

Where $N$ is the number of hops on the packet's path. For the packet to meet it's deadline, the end-to-end delay must not exceed its relative deadline, $D_r$. Therefore,

$$\sum_{x=1}^{N} \frac{VQ_x(1 - VQ_x/2)}{1 - VQ_x} D_{max} \leq D_r \quad (9)$$

Rearranging the terms, we get:

$$\sum_{x=1}^{N} \frac{VQ_x(1 - VQ_x/2)}{1 - VQ_x} \leq D_r / D_{max} \quad (10)$$

To obtain a sufficient bound, the ratio of the deadline of packet $P_a$ to that of a higher priority packet that delays its transmission, $D_r/D_{max}$ must be minimized over all possible cases. Let us denote the minimum of this ratio for a given fixed priority scheduling policy $\delta$. Deadline monotonic algorithm assigns priority in the order of decreasing relative deadlines. Thus we define, $\delta = 1$ for deadline monotonic scheduling algorithm. Clearly, this ratio can not be larger than 1. Therefore, deadline monotonic scheduling algorithm gives the maximal bound:

$$\sum_{x=1}^{N} \frac{VQ_x(1 - VQ_x/2)}{1 - VQ_x} \leq 1 \quad (11)$$

4.3) EDF Scheduling

We use Theorem 4.2 to derive schedulability bound on neighborhood utilization for EDF. Similar to the derivation for fixed priority scheduling algorithms presented in the previous section, we consider transmission of a packet along some path whose nodes are labeled $1 \ldots N$. Since $VQ_x$ is the neighborhood utilization of each node $x$ on this path, the total utilization on the path is $\sum_{x=1}^{N} VQ_x$. From Theorem 4.2 all packets on this path are schedulable provided

$$\sum_{x=1}^{N} VQ_x \leq 1 \quad (12)$$

Theorem 4.2 ($U_{Bound_4}$), $U = \sum_{i=1}^{n} u_i \leq 1$ that is $U_{Bound_4} = 1$, is a sufficient condition for schedulability of jobs $J1, \ldots\ldots, Jn$ in a pipeline using EDF if the decomposition of the execution times of the jobs is unknown.

**5) Capacity Limits**

In this section, we present the analysis of real-time capacity limits for Deadline Monotonic and Earliest Deadline First scheduling algorithms.

5.1) Load balanced traffic

Load balanced traffic refers to the traffic pattern where every node has the same node utilization.

Equivalently every neighborhood utilization becomes the same. More precisely, every neighborhood utilization equals the lowest bound of all neighborhood utilizations, determined by the upper bound on neighborhood size $u$. We denote this common bound on neighborhood utilization by $VQ_x$. Let N be the upper bound on the path length. Therefore, from (11),

$$\sum_{x=1}^{N} \frac{VQ_x(1-VQ_x/2)}{1-VQ_x} \leq 1$$

$$\Rightarrow \sum_{x=1}^{N} \frac{VQ_x(1-VQ_x/2)}{1-VQ_x} \leq 1/N \quad (13)$$

Solving for $VQ_x$ we get

$$VQ_x = 1/N + 1 - \sqrt{1/N^2 + 1} \quad (14)$$

From (5)

$$RTCC_{FP} = \frac{nB}{u\alpha}\left(1/N + 1 - \sqrt{1/N^2 + 1}\right)$$
(15)

Where n is the number of nodes in the network.

Now we obtain the real time capacity for EDF from (11)

$$\sum_{x=1}^{N} VQ_x \leq 1 \quad \Rightarrow VQ_x \leq 1/N$$
(16)

$$RTCC_{EDF} = \frac{nB}{uN\alpha}$$
(17)

Figure 1 shows a comparison of the real-time capacity limit expressions for DM and EDF. The figure shows that the real-time capacity limit for DM approaches that for EDF as the path length increases. This behavior of the curves can be explained as follows: as paths get larger $1/N^2 \to 0$. Therefore $RTCC_{FP} \to \frac{nB}{uN\alpha} = RTCC_{EDF}$.

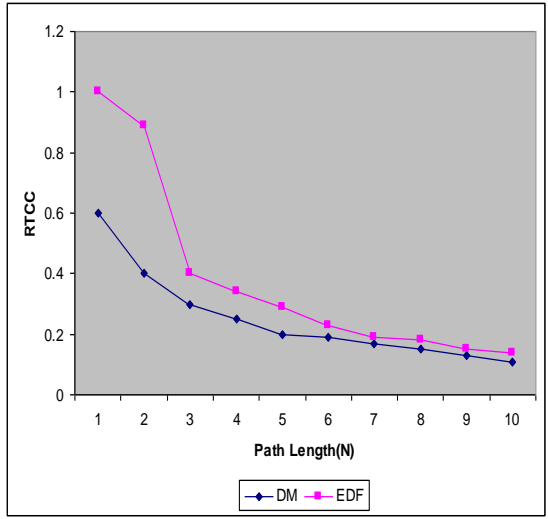
Fig 1 Real time communication capacity

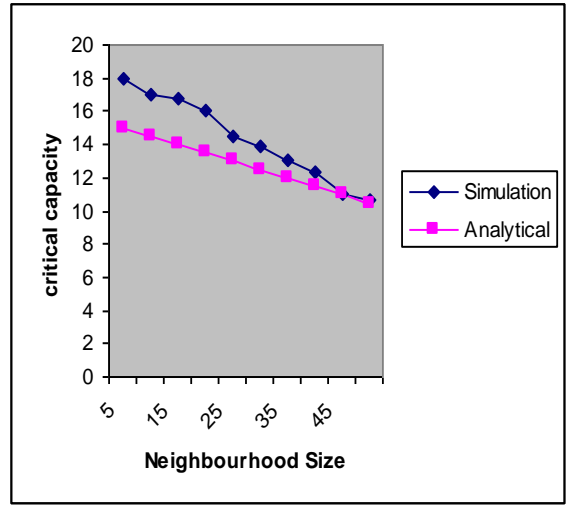
Fig 2 Effect of radio radius, 800 nodes

### 6) Convergecast traffic

Convergecast refers to the traffic pattern in which nodes transmit data to a common data aggregation point or base-station (Figure 3). There may be a number of data aggregation points distributed in the network. In such a case, the traffic is destined to the nearest data aggregation point. The data aggregation points form the bottleneck. The nodes in the neighborhood of the data aggregation points are loaded the most. In the following, we derive an approximate expression for real time capacity limit for convergecast.

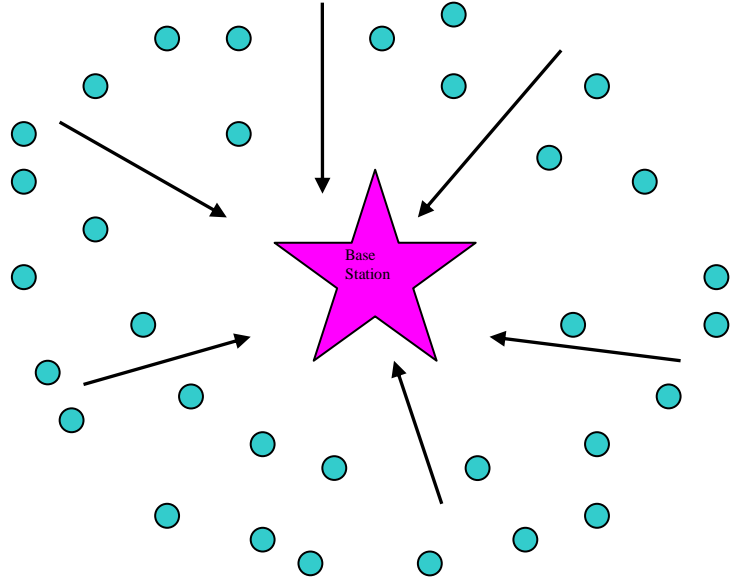

Figure 3 Convergrcast traffic

Let N be the number of data aggregation points in the network. We consider the traffic sent to an arbitrary data aggregation point d. Let $K_D$ be largest path length of the packets destined to d. Since all traffic ends up at the data aggregation point d, at steady state, the total amount of traffic generated for d must equal the total amount that can be delivered to d. Let $\frac{m}{\Pi R^2}$ be the average density of nodes in the area of one radio range R. The number of nodes located x hops away from the data aggregation point is approximated by the product of the area of the ring $\Pi x^2 R^2 - \Pi(x-1)^2 R^2$ and the node density $\frac{m}{\Pi R^2}$ which is equal to $(2x-1)m$ nodes. Therefore, the average node utilization, and hence neighborhood utilization, decreases with distance from the aggregation point. If we denote the utilization at the data aggregation point to be $D$, then the utilization of the xth hop nodes is $D_x = D/(2x-1)m$. Therefore, from the feasibility condition for DM is

$$\sum_{x=1}^{K_d} \frac{D_x(1-D_x/2)}{1-D_x} \leq 1$$

$$\Rightarrow \sum_{x=1}^{K_d} \frac{\frac{D}{(2x-1)m}(1-\frac{D}{(2x-1)2x})}{1-\frac{D}{(2x-1)m}} \leq 1$$

$$\Rightarrow \sum_{x=1}^{K_d} \left\{ \frac{D}{(2x-1)m} \frac{2(2x-1)m-D}{2(2x-1)m-D} \right\} \leq 1$$

$$\Rightarrow \sum_{x=1}^{K_d} \left\{ \frac{D}{2(2x-1)m} \left(1 + \frac{(2x-1)m}{(2x-1)m-D}\right) \right\} \leq 1 \quad (18)$$

Inequality (18) is nonlinear in D. We present a closed form expression that approximates the exact solution below. From (5),

$$RTCC_{FP} = NBK_d D / \alpha \qquad (19)$$

where D is solution of (18).
Now we obtain the real-time capacity for EDF. From (11),

$$\sum_{x=1}^{K_d} VQ_x \leq 1$$

$$\Rightarrow \quad D/m \sum_{x=1}^{K_d} \frac{1}{2x-1} \leq 1 \qquad (20)$$

To obtain the sum $\sum_{x=1}^{K_d} \frac{1}{2x-1}$, we note that the Euler's constant $\gamma (\cong 0.587)$ is given by

$$\gamma = 1 + 1/2 + 1/3 + \ldots + 1/x - \log x \qquad (21)$$

Where the series approximation gets better as $x \to \infty$. Let $S = \sum_{x=1}^{K_d} \frac{1}{2x-1}$, and let $S' = \sum_{x=1}^{K_d} 1/2x$. From $S' = 1/2(\gamma + \log K_d)$. Furthermore,

$$S + S' = 1 + 1/2 + 1/3 + \ldots + 1/2K_d$$
$$= \gamma + \log 2k_d \qquad (22)$$

Therefore,

$$S = \gamma/2 + \log 2K_d - 1/2 \log 2k_d$$
$$= \gamma/2 + \log 2 + 1/2 \log k_d$$
$$\cong 1 + 0.5 \log K_d \qquad (23)$$

We get D by plugging (23) into (20) as:

$$D = \frac{m}{1 + 0.5 \log K_d} \qquad (24)$$

From (5)

$$RTCC_{EDF} = \frac{NBK_d}{\alpha (1 + 0.5 \log K_d)} \qquad (25)$$

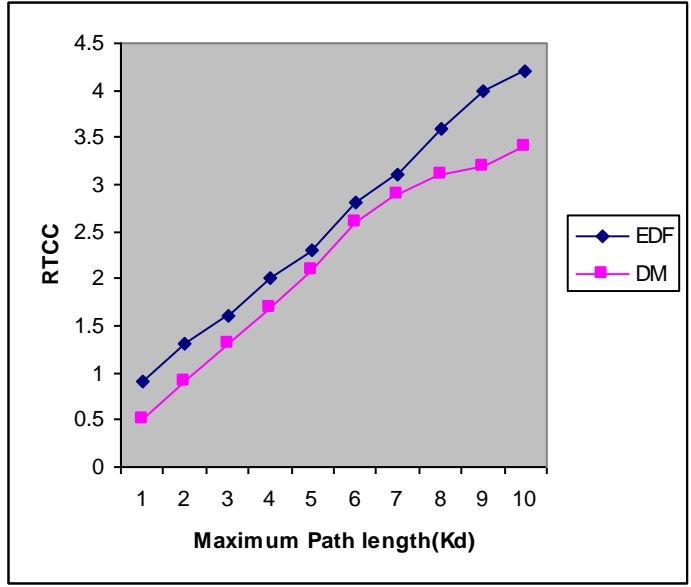

Figure 4 Real time communication capacity for convergecast traffic.

Figure 4 presents a graph of the real-time capacity limit for DM and EDF scheduling algorithms. The values were obtained using numerical solution of the capacity expressions derived above. The plot shows that the relative difference in the capacity expressions get diminished as Kd increases. This can be reasoned by noting that in (18), the second term can be approximated by unity, given that D < 1. Upon this approximation, (18) becomes same as the bound on D for EDF

$$RTCC_{FP} \approx \frac{NBK_d}{\alpha(1+0.5\log K_d)} \quad (26)$$

6.1) Convergecast vs. Load-balanced Traffic

In the following, we compare the real-time capacity limits obtained for load balanced traffic and convergecast traffic. For the purpose of comparison, we use the expressions for EDF since the approximate closed form expressions for DM resemble that for EDF. Let $RTCC^{BL}$ and $RTCC^{CC}$ be the real-time capacity limits for the load balanced and convergecast cases respectively. Then, for the similar path lengths, namely $K_d = K_d'$ we have

$$\frac{RTCC^{BL}}{RTCC^{CC}} = \frac{n(1+0.5\log K_d)}{NmK_d^2} \quad (27)$$

Since $mK_d^2$ is approximately the number of nodes inside one aggregation point domain, $mK_d^2 \cong n/N$. Therefore

$$\frac{RTCC^{BL}}{RTCC^{CC}} \cong 1+0.5\log K_d \quad (28)$$

This ratio becomes 1 when $K_d = 1$. This is to be expected since each transmission excludes $m-1$ nodes of the neighborhood in both traffic topologies. For larger values of Kd, the capacity limit for load balanced case is larger than that for the convergecast since the absence of bottleneck in the earlier case allows the reception of larger amount of traffic at the destination per unit time. The gain, however, is only logarithmic. The reason for this phenomenon is that for the load balanced traffic, the intersection of different flows, and hence interference grows with the path size. But, for the convergecast traffic, the traffic flow is radial, and hence the interference is not much of an issue at larger hop distances from data aggregation points.

### 7) Evaluation

We implemented a simulator J-sim to evaluate the pessimism in the capacity expressions. The simulator constructs a network of sensor nodes of a user-specified size in a perturbed grid structure. The radio layer is implemented as a simplified disk model of a specified radius (range). The sinks are distributed uniformly across the network. We generated traffic at each non-sink node such that each packet was assigned a deadline at random from a preselected set. All packets were sent to their nearest sink. Packet contention was resolved in priority order. Only those nodes were allowed to transmit who were not within the radio range of another node that was already scheduled to receive a transmission. Ties between simultaneously arriving same priority packets were broken at random. We implemented a shortest path routing scheme in which the neighboring node nearest to the sink was chosen as the next hop. If this node was blocked due to another transmission, the packet was not scheduled until that transmission was over. The MAC layer implements deadline monotonic prioritization for medium arbitration. All packets were checked for deadline misses. All packets were checked for deadline misses. If there was a miss, the actual capacity consumption of all in-transit traffic was computed by multiplying each in transit packet by the traversed hop count and normalizing by the end-to-end deadline. Each run was repeated 50 times with different randomized workloads. The minimum capacity consumption at which a deadline miss occurred was recorded, and is shown on the plots as critical capacity.

Figures 2 show the effect of increasing the radio radius, shown on the top horizontal axis, on real-time capacity in a network of 800 nodes. Observe that increasing the radio radius also increases the neighborhood size (i.e., the number of nodes within the radio range), shown on the bottom horizontal axis. The number of sinks was kept at 12.

Figures 5 and 6 repeat the experiments for networks of 800 and 1600 nodes respectively, this time varying the number of sinks. The radio range is kept constant at a neighborhood size of 12 nodes. As before, a very close match is observed between simulation and analysis. Capacity grows with the number of sinks because data collection bottlenecks are alleviated.

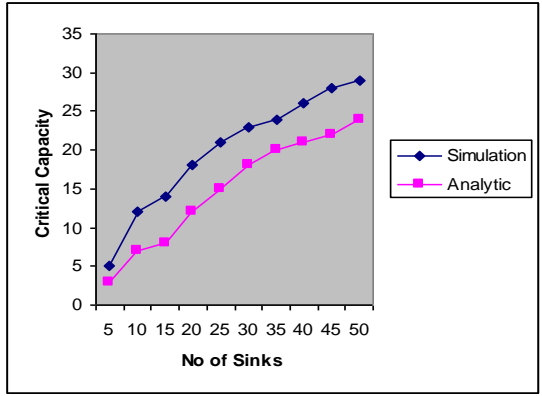

Figure 5 Effect of the number of sinks, 800 nodes

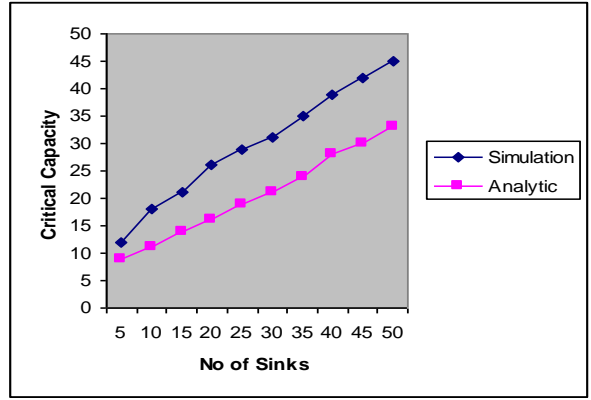

Figure 6 Effect of the number of sinks, 1600 nodes

Finally, Figure 7 shows the sharp increase in the miss ratio in a network of 800 nodes that occurs when capacity is exceeded. In this curve, the network workload is increased past the capacity bound. The miss ratio is then plotted against the capacity requirements of the workload shown on the horizontal axis. Each point in the figure corresponds to a single experiment. Two sets of data points are shown for two different radio ranges that correspond to neighborhoods of 12 nodes and 24 nodes respectively.

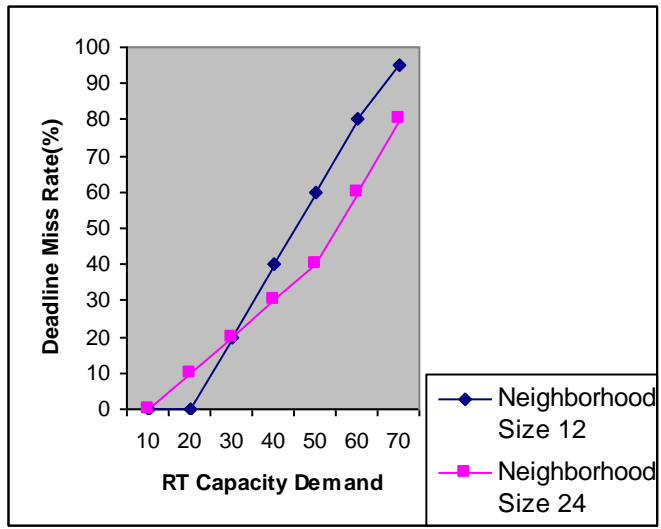

Figure 7 Miss rate as a function of real time capacity demand

## 8) Conclusions

In this paper we presented real-time capacity limit expressions under the sufficient schedulability condition. We derived limiting expressions for large networks. We saw that under such limiting circumstances, the real-time capacity limits for deadline monotonic and earliest deadline first scheduling algorithms tend to agree on a common value. We showed that for convergecast, the real-time capacity increases as the square root of the number of nodes. This implies existence of diminishing return and hence favors multiple smaller networks as opposed to a large one. We find that the real-time capacity obtained analytically agrees with that obtained using simulation within 30-40% in general. The agreement gets better as the effective density increases.

## References

[1] P. Gupta and P. R. Kumar. The capacity of wireless networks. Information Theory, IEEE Transactions on, 46(2):388–404, March 2000.

[2] Tarek Abdelzaher, K. Shashi Prabh, and Raghu Kiran. On real-time capacity limits of multihop wireless sensor networks. In Proceedings of the 25th IEEE Real-Time Systems Symposium (RTSS'04). IEEE Computer Society Press, Los Alamitos, CA, 2004.

[3] Ting Yan, Tian He, and John A. Stankovic. Differentiated surveillance for sensor networks. In SenSys '03: Proceedings of the 1st international conference on Embedded networked sensor systems, pages 51–62, New York, NY, USA, 2003. ACM Press.

[4] Qing Cao, Ting Yan, John Stankovic, and Tarek Abdelzaher. Analysis of target detection performance for wireless sensor networks. In Distributed Computing in Sensor Systems, First IEEE International Conference, Proc. of (DCOSS 2005), pages 276–292. Springer, 2005.

[5] Ting Yan. Analysis Approaches for Predicting Performance of Wireless Sensor Networks. PhD thesis, University of Virginia, 2006.

[6] Jean-Yves Le Boudec and Patrick Thiran. Network calculus: a theory of deterministic queuing systems for the Internet. Springer-Verlag New York, Inc., New York, NY, USA, 2001.

[7] K. Shashi Prabh and Tarek F. Abdelzaher. Energy conserving data cache placement in sensor networks. The ACM Transactions on Sensor Networks, 1(2):178–203, November 2005.

[8] Anis Koubaa, Mario Alves, and Eduardo Tovar. Modeling and worst-case dimensioning of cluster-tree wireless sensor networks. In Proceedings of the 27[th] IEEE Real-Time Systems Symposium (RTSS'06), pages 412–421, Los Alamitos, CA, USA, 2006. IEEE Computer Society.

[9] Tarek Abdelzaher, Gautam Thaker, and Patrick Lardieri. A feasible region for meeting aperiodic end-to-end deadlines in resource pipelines. In Proceedings of the 24th International Conference on Distributed Computing Systems(ICDCS'04), pages 436–445. IEEE Computer Society Press, Los Alamitos, CA, 2004.

[10] T. P. Baker. Multiprocessor EDF and deadline monotonic schedulability analysis. In Proc. of the 24th IEEE Real-Time Systems Symposium (RTSS '03), pages 120–129. IEEE Computer Society Press, Los Alamitos, CA, 2003.